\documentclass[ prd,  twocolumn,floatfix,showpacs,showkeys]{revtex4-1}
\usepackage{latexsym,amssymb,amsmath,amsfonts}
\usepackage[mathscr]{euscript}
\usepackage{graphicx}
\usepackage{bm}
\usepackage{multirow}
\usepackage{setspace}
\newcommand{\beq}{\begin{equation}}
\newcommand{\eeq}{\end{equation}}
\newcommand{\bc}{\begin{center}}
\newcommand{\ec}{\end{center}}
\newcommand{\eeqa}{\end{eqnarray}}
\newcommand{\beqa}{\begin{eqnarray}}
\newcommand{\no}{\noindent}
\newcommand{\pa}{\partial}

\newcommand{\ra}{\rightarrow}
\newcommand{\na}{\nabla}

\newcommand{\al}{\alpha}
\newcommand{\be}{\beta}
\newcommand{\ga}{\gamma}
\newcommand{\Ga}{\Gamma}
\newcommand{\de}{\delta}
\newcommand{\De}{\Delta}
\newcommand{\ep}{\epsilon}
\newcommand{\ze}{\zeta}
\newcommand{\et}{\eta}
\newcommand{\ka}{\kappa}
\newcommand{\la}{\lambda}

\newcommand{\si}{\sigma}

\newcommand{\ph}{\phi}

\newcommand{\ps}{\psi}

\newcommand{\om}{\omega}
\newcommand{\ed}{\end{document} }

\begin{document}

\title{Gravity with spin and electromagnetism with a nonsymmetric metric tensor}
\author{Richard T. Hammond}
\email{rhammond@email.unc.edu }
\affiliation{
University of North Carolina at Chapel Hill\\
Chapel Hill, North Carolina and\\
Army Research Office\\
Research Triangle Park, North Carolina}

\date{\today}


\begin{abstract}
It is shown the antisymmetric part of the metric tensor is the potential for the spin field. Various metricity conditions are discussed and comparisons are made to other theories, including Einstein's. It is shown in the weak field limit the theory reduces to one with a symmetric metric tensor and totally antisymmetric torsion. It is also shown, due to gauge invariance, the electromagnetic field must be present.
\end{abstract}

\maketitle

\section{introduction}

One of the strongest motivations for developing a theory with a nonsymmetric metric tensor comes from the nonsymmetric energy momentum tensor, but first,  let us address a common proof, appearing in many textbooks, that the energy momentum tensor must be symmetric. These ``proofs" assume  the angular momentum is orbital angular momentum, so the contribution from a small region is the density times the volume times the radius. As the volume goes to zero so does the contribution to the angular momentum, which, as the proof shows, yields a symmetric energy momentum tensor. However, if the angular momentum is intrinsic spin the argument fails, as discussed in detail by Sciami.\cite{sciama} In fact, Papapetrou showed the antisymmetric part of the energy momentum tensor is related to  spin (by spin I will always mean intrinsic spin), \cite{bel},\cite{pap} and so we are naturally led to consider a nonsymmetric  metric tensor if spin is included.

The exception to this argument arises from gravitation with a symmetric metric tensor with torsion. In this case, as is known, the torsion can describe spin. However, the torsion  arises from an antisymmetric source tensor and, as we will see below for torsion of the string theory type,  the source of torsion turns out to be the antisymmetric part of the metric tensor. Thus, we end up with a nonsymmetric energy momentum tensor.

These arguments  also tell us what the physical interpretation of the nonsymmetric part of the metric tensor is, it is related to spin. However, all this assumes the source has intrinsic spin. For bodies like the Earth or an apple the spins may add to zero. However, the source used for composite bodies is always some averaged out model of the true source, the elementary particles that make up matter. To have a realistic theory, it must be formulated in terms of a source which is an elementary particle (macroscopic averaging may be done later), and each elementary particle has intrinsic spin. The electron,  quarks, and the electron neutrino, and the two other families each have spin. Exempted here are the two exchange particles with zero spin; the Higgs and the quantum of torsion. With this proviso, since particles have spin, the energy momentum tensor cannot be symmetric and so, we take the metric tensor to be nonsymmetric.

More modern developments in physics, after the Einstein era, make the nonsymmetric metric tensor more cogent than ever. Anti-commuting variables, non-commutative geometry and superspace can be related to the antisymmetric part of the metric tensor. For example, for anti-commuting Grassmann variables $\theta^\mu$, which satisfy $\{d\theta^\mu,d\theta^\nu\}=0$, the line element can be generalized to include $\ph_{\mu\nu}d\theta^\mu d\theta^\nu$ where $\ph_{\mu\nu}$ is the antisymmetric part of the metric tenor. In the metricity conditions (discussed fully below), the antisymmetric part of the metric tensor appears in the skew symmetric derivative, $\ph_{\mu\nu,\si}+\ph_{\si\mu,\nu}+\ph_{\nu\si,\mu}$. The temptation to set this to zero may have fueled attempts to interpret $\ph_{\mu\nu}$ as being related to the electromagnetic field, but today we see this combination is used to define torsion (as discussed below) both from generalized gravity\cite{tg} and string theory.\cite{st} 

The older work is well known. Soon after Einstein published his theory of general relativity he, and others, began serious investigations into the physical theory based on a non-symmetric metric tensor $g_{\mu\nu}$ \cite{e1},\cite{e2},\cite{e3}, a detailed review may be found in the literature.\cite{goen} For almost 100 years this search led to many dead-ends and disappointing quagmires. 
Although early investigations of Einstein sought to develop a fully geometric theory in which the energy momentum tensor was of a geometric origin, most of his efforts sought to find a unified theory of gravitation and electromagnetism. Various researchers picked up the torch, including Schr\"odinger,\cite{s} but finally the light dimmed as it became evident these dark paths led nowhere.

Two notable approaches followed. One is the work of Sciama,\cite{sciama} who showed the antisymmetric part of the metric tensor is associated with intrinsic spin. This followed work by Papapetrou and Belinfante
who considered a nonsymmetric energy momentum tensor.\cite{pap} Sciama adopted Einstein's variational approach but did not associate the theory with electromagnetism. Soon after, however, Sciama adopted a symmetric metric tensor but considered spacetime to be endowed with torsion. This notion of classical spin in general relativity with a symmetric metric tensor was carried forward by Hehl and his collaborators in the 1970s and later others.\cite{hehl} \cite{hamrev} \cite{shap} The other well-known approach is due to Moffat.\cite{moffat} He adopted Einstein's equations with a nonsymmetric metric tensor, but he eschewed the notion of electromagnetism, taking the theory to be one of gravitation only.
It was shown if the metric tensor was expanded about a background metric, the antisymmetric potential appears without a derivative. This non gauge invariant term gave rise to ghosts, but they were removed by adding cosmological type terms.\cite{dam} The revenant idea of the antisymmetric part of the metric tensor being associated with spin was described more recently.\cite{ham1}

One of the problems with using a nonsymmetric metric is the proliferation of tensors, such as in the definition of the covariant derivative, the curvature tensor, and its contractions that form scalars to be used in the Lagrangian. Einstein used transposition invariance as a guiding principle. This is not a physics principle, unless one interprets it as coming from a charge invariance,\cite{mc}, but even if this fragile notion stood up, it is based on the idea the theory involves electromagnetism. 
Even recently, for example, the notion that the antisymmetric part of the metric tensor is associated with nuclear forces has been considered.\cite{yildez}

In the current paper 
it is shown the non-symmetric part of the metric tensor is associated with spin.
With this,
physical principles are used to choose the appropriate terms so that the proliferation of tensors is not an issue.

\subsection{\label{previous} previous related work}

It will be shown that the linearized version of the nonsymmetric theory developed below  reduces to a well studied
 theory  with a symmetric metric tensor and torsion of the string theory type, i.e.,
 
 \beq\label{tordef1}
 S_{\mu\nu\si}\equiv \ps_{[\mu\nu,\si]}= \frac13(\ps_{\mu\nu,\si}+\ps_{\si\mu,\nu}+\ps_{\nu\si,\mu})
 \eeq
where
\beq
S_{\mu\nu}^{\ \  \si}=\frac12\left(\Gamma_{\mu\nu}^{\ \ \si}-\Gamma_{\nu\mu}^{\ \ \si}
\right)
\eeq
where $\Gamma_{\mu\nu}^{\ \ \si}$ is the affine connection
and where $\ps_{\mu\nu}$ is the antisymmetric torsion potential. The geometrical part of the action is simply 

\beq\label{r}
I=\int\sqrt{-g}d^4xR
\eeq
where $R$ is the curvature scalar and, in that theory,  variations are taken with respect to a general potential defined by $\Phi_{\mu\nu}\equiv g_{\mu\nu}+\psi_{\mu\nu}$ where $g_{\mu\nu}$ is the symmetric metric tensor.\cite{hamrev} This theory is very successful. One of the most important results  is that the correct conservation law of total angular momentum is obtained if the source of torsion is intrinsic spin. From this viewpoint, it was argued that torsion must exist. A phenomenological source was constructed and the weak field limit of the torsion was found. This coincided exactly with the low energy limit of the Dirac equation in spacetime  with torsion, which is a very important confirmation of the theory. It was then shown strings are a natural source for the material action, and it was further shown how the correct equations of motion arise from strings. However, the notion of a general potential  $\Phi_{\mu\nu}$ was not a geometrical foundation. It was simply a tidy way of deriving the field equations. Variations with respect to the general potential are identical to independent variations of the metric tensor and torsion potential.
However, it was suggestive. It certainly invites the question, can the theory be formulated in terms of a nonsymmetric metric tensor? In addition, it is known the skew symmetric derivative appearing in (\ref{tordef1}) arises naturally in the nonsymmetric theory when metricity conditions are used. Instead of thinking this is related to electromagnetism, as Einstein did, it is more natural that this may be related to spin, as will be seen in detail below. 

 This notion gives us very useful direction, and especially what to expect in the weak field limit. Also, the notion that the (dimensionless) antisymmetric part of the metric potential acts like a potential for the field coincides with the idea that the symmetric part is like the potential of the gravitational field. Finally, one of the original complaints about the nonsymmetric theory was that the symmetric part and antisymmetric part transform separately, which was considered bad. However, the gravitational field is not the source of spin and the spin field is not the source of gravity, and no coordinate transformation should make that so, so this bemoaned separation is actually a good thing. Finally, as further motivation, it was recently shown that the spin of the electromagnetic field gives rise to torsion if the energy momentum tensor is nonsymmetric.\cite{hamx} A nonsymmetric energy momentum tensor requires the existence of a nonsymmetric metric tensor, and in this case it is related to spin.

The field equations for the theory with a symmetric metric tensor from (\ref{r}) with a material action added,  and
with (\ref{tordef1}), are given by 

\beq\label{fe}
G^{\mu \nu}  - 3S^{\mu \nu \sigma}_{\ \ \ \ ; \sigma}
-2S^{\mu}_{\ \alpha \beta}S^{\nu \alpha \beta} = kT^{\mu \nu}
\eeq
where $k=8\pi G$ and $G^{\mu\nu}$ is the Einstein tensor in U4 spacetime (spacetime with torsion) and the semicolon is the Levi-Civita covariant derivative.
The symmetric part yields the gravitational field equations,
\beq\label{gfe}
G^{(\mu \nu)} -2S^{\mu}_{\ \alpha \beta}S^{\nu \alpha \beta} =kT^{(\mu\nu)}
,\eeq
or, in terms of the Riemannian Einstein tensor $ ^oG^{\mu\nu}$,

\beq\label{gfer}
 ^oG^{\mu \nu} -3S^{\mu}_{\ \alpha \beta}S^{\nu \alpha \beta}
 +\frac12 S^{ \alpha \beta\si}S_{ \alpha \beta\si}
 =kT^{(\mu\nu)}
,\eeq
and the antisymmetric part are the torsional field equations,

\beq\label{tfe}
S^{\mu \nu \sigma}_{\ \ \ \ ;\sigma} =-kj^{\mu \nu}
\eeq
where $j^{\mu\nu}\equiv (1/2)T^{[\mu\nu]}$.

It should also be noted the curvature scalar, $R$, of space-time with torsion given by (\ref{tordef1}) 
can be written in terms of the Riemannian scalar $ ^oR$ according to

\beq\label{rplus}
R=\ ^oR-S_{\al\be\si}S^{\al\be\si}
.\eeq
It is important to note this is the same Lagrangian in the low energy limit of string theory without the scalar field (or constant scalar field.) However, the scalar field arises naturally, and cannot be zero, when this is coupled to the Dirac equation.\cite{ham2}

\section{affine geometry}
We begin by looking at a spacetime endowed with an affine connection, more details may be found in the literature.\cite{ton} At this point there is no concept of a metric tensor or distance. In an affinely connected space we have parallel transport entering into the covariant derivative according to, for any vector $A^\si$,

\beq\label{-}
\na_{\underset -\mu} A^\si =
A^\si_{\ ,\mu}+\Gamma_{\mu \nu}^{\ \ \si}A^\nu
\eeq
or

\beq\label{+}
\na_{\underset +\mu}  A^\si =
A^\si_{\ ,\mu}+\Gamma_{\nu\mu}^{\ \ \si}A^\nu
\eeq

\no where the plus or minus sign underneath the index labels these two definitions.
Since
 
\beq
\na_{\underset -\mu }A^\si -\na_{\underset +\mu }A^\si =
(\Gamma_{\mu \nu}^{\ \ \si}-\Gamma_{\nu\mu}^{\ \ \si})A^\nu
\equiv 2S_{\mu \nu}^{\ \ \si}A^\nu
,\eeq
these definitions are different unless the torsion tensor, $S_{\mu \nu}^{\ \ \si}$, vanishes.
The first, (\ref{-}), will be adopted here. Starting with the second instead, the final results are equivalent.

The curvature tensor is defined according to

\beq\label{loop}
\De A^\si=\frac1 2A^\nu R_{\be\mu\nu}^{\ \ \ \ \si}\oint \xi^\mu dx^\be
\eeq
\no where $A^\si$ is parallel transported along the small closed curved according to (\ref{-}), and is given by

\beq\label{curveten}
R_{\be\mu\nu}^{\ \ \ \ \si}=
\Ga_{\mu\nu ,\be}^{\ \ \si}-\Ga_{\be\nu ,\mu}^{\ \ \si}
+\Ga_{\be\ph}^{\ \ \ \si}\Ga_{\mu\nu}^{\ \ \ \ph}
-\Ga_{\mu\ph}^{\ \ \ \si}\Ga_{\be\nu}^{\ \ \ \ph}
\eeq
and
contracting gives the Ricci tensor,

\beq\label{ricci}
R_{\mu\nu}=R_{\si\mu\nu}^{\ \ \ \ \si}\\ 
,\eeq

\no which is not symmetric. 
There is also the segmental curvature tensor 
$V_{\mu\nu}$ defined by
\beq
V_{\mu\nu}\equiv R_{\mu\nu\si}^{\ \ \ \ \si}
=\Ga_{\nu,\mu}-\Ga_{\mu,\nu}
\eeq
where $\Ga_\nu\equiv\Ga_{\nu\si}^{\ \ \si}.$
\no The antisymmetric part of the Ricci tensor is given by

\beq\label{aricci}
R_{[\mu\nu]}=-\frac12V_{\mu\nu}+\na_\si S_{\mu\nu}^{\ \ \si}
+3(\na_{[\mu}S_{\nu ]}
+S_{\mu\nu}^{\ \ \si}S_\si )
\eeq
where the brackets imply taking the antisymmetric part and $S^\nu=S^{\nu\si}_{\ \ \si}$ is called the torsion trace or
the torsion vector. 
The Bianchi identities are given by

\begin{equation}\label{bi}
\nabla_{\nu}G^{\mu \nu} = 2S^{\mu \alpha \beta}R_{\beta \alpha}
-S_{\alpha \beta \si}R^{\mu \si \beta \alpha}
.\end{equation}

\section{metrical space}

The metric tensor may be written in terms of its symmetric part, $\ga_{\mu\nu}$ and its antisymmetric part $\ph_{\mu\nu}$,

\beq\label{mt}
g_{\mu\nu}=\ga_{\mu\nu}+\ph_{\mu\nu}
.\eeq
For any $n\mbox{x}n$ array, $N$, we define the inverse according to  $N^{\mu\theta}N_{\nu\theta}=\de^\mu_\nu=
N^{\theta\mu}N_{\theta\nu}$, so for example, we define $g^{\mu\nu}$ so that 

\beq\label{gde}
g^{\mu\theta}g_{\nu\theta}=\de^\mu_\nu
\eeq and 
\beq\label{gade}
\ga^{\mu\theta}\ga_{\nu\theta}=\de^\mu_\nu
.\eeq 
However, we shall define the contravariant form of the metric tensor in an unconventional form, i.e.

\beq\label{mtcont}
g^{\mu\nu}=\ga^{\mu\nu}+f^{\mu\nu}
.\eeq
In this form, since both (\ref{gde}) and (\ref{gade}) hold, $f^{\mu\nu}$ is determined, and is given by

\beq\label{f}
f^{\mu\nu}=-\ga^{\nu\al}g^{\mu\be} \ph_{\al\be}
.\eeq
This last form involves the (inverse) metric tensor again on the right side, but that is defined in terms of  (\ref{gde}) so this is well defined. It is most useful in linearizing the theory. In fact, an iterative solution may be written by using (\ref{mtcont}) in (\ref{f}), giving

\beq\label{f2}
f^{\mu\nu}=-\ga^{\nu\al}\ga^{\mu\be} \ph_{\al\be}-\ga^{\nu\al}f^{\mu\be} \ph_{\al\be}+\ldots \ 
.\eeq

Unlike GR, there is no unique way of raising and lowering indices. If we consider the vector $A_\si$, then there are two different contravariant forms, $g^{\si\mu}A_\mu$ and $g^{\mu\si}A_\mu$.  Each of these is a bonafide tensor, but we take the second as the definition of $A^\si$.  To lower an index we assume $A_\mu
=A^\si g_{\mu\si}$.
This allows us to consistently raise and lower indices.

The next decision is the most important considering the resulting field equations. It is, whether to fix a condition between the metric tensor and the affine connection and then consider variations with respect to the metric tensor and its derivates, (this is  called the second order formalism), or to consider the metric tensor and the affine connection independent and consider variations with respect to each. This is called the Palatini method. In GR they give the same result, but in general they do not.
In order to see the details we consider nonmetricity in some detail.

\subsection{meaning of metricity}
The notion of setting the covariant derivative of the metric tensor to zero  is rooted in physics.
In GR it guarantees the scalar invariant of two vectors parallel transported along a curve remains constant, which has the immediate ramification that the length of a vector does not change upon parallel transport.  For example, Einstein used this idea to show Weyl's unified theory was unphysical. Here it is assumed that metricity holds, but there is more to the story.

Let us consider the  vector $A^\mu$  undergoing parallel transport along an infinitesimal $dx^\si$. Define the scalar product as $P= A^\mu A^\nu g_{\mu\nu}$. Then, according to (\ref{-}),
\beq\label{dp}
	dP=-A^\mu A^\nu\na_\si g_{{\mu}{\nu}}dx^\si 
\eeq
and one may hasten to the conclusion that if the scalar product remains constant then $\na_\si g_{\mu\nu}=0$. However, if we adopt a different definition of covariant differentiation then (\ref{dp}) will change.
More importantly,  the entire notion of metricity has no real foundation when the metric tensor is not symmetric. This is true for two reasons. From (\ref{dp}) we see the antisymmetric part of the metric tensor drops away and therefore the covariant derivative of $\ph_{\mu\nu}$ is not specified. The second has its roots in physics. Although the notion of parallel transport is essential in developing the theory of curved space, it does not represent the physical motion of a particle. The equation of motion must be derived from the Bianchi identities, and it is known the motion is generally not that of parallel transport. Thus parallel transport along a curve does not represent
a natural motion in curved space, and could only be achieved with additional, nongravitational, forces.

It is stressed that once we adopt the definition of covariant differentiation given by (\ref{-}), the covariant derivative of a second order (and higher) tensor is already defined. For example, with (\ref{-}), the covariant derivative of a second rank tensor $T_{\mu\nu}$ is

\beq\label{--}
		\na_\si T_{\mu\nu}=T_{\mu\nu},_\si - \Gamma_{\si\mu}^{\ \ \ \theta}T_{\theta\nu}
-\Gamma_{\si\nu}^{\ \ \theta}T_{\mu\theta}=0
.\eeq

However Einstein adopted the condition,

\beq\label{ein}
\na_\si g_{{\underset +\mu }{\underset -\nu}}=g_{\mu\nu},_\si-\Gamma_{\mu\si}^{\ \ \theta}g_{\theta\nu}
-\Gamma_{\si\nu}^{\ \ \theta}g_{\mu\theta}=0
,\eeq
which is not the same as setting $\na_\si g_{\mu\nu}=0$. This issue can be resolved by adopting the metricity condition of the following form.

\beq\label{q}
\na_\si g_{\mu\nu}
\equiv
\na_\si g_{{\underset-\mu}{\underset-\nu}}
=q_{\si\mu\nu}
\eeq
where  $q_{\si\mu\nu}$ is the nonmetricity tensor. With $q_{\si\mu\nu}=-2S_{\si\mu\nu}, $ (\ref{--}) and (\ref{ein}) are the same (we may assume the plus minus derivative used by Einstein and others is related to the choice of the metricity tensor). But why choose this?

It turns out we can choose the nonmetricity on physical grounds. To see this we will work out the various cases of metricity.  Following this notion, let us work out the natural possibilities according to the possible choices of

\no\beq\label{mm}
1.\ \ \
		\na_\si g_ {\underset -\mu \underset-\nu}=g_{\mu\nu},_\si-\Gamma_{\si\mu}^{\ \ \theta}g_{\theta\nu}
-\Gamma_{\si\nu}^{\ \ \theta}g_{\mu\theta}=0
\eeq

\beq\label{-+}
2.\ \ \ 
\na_\si g_{{\underset -\mu }{\underset +\nu}}=g_{\mu\nu},_\si-\Gamma_{\si\mu}^{\ \ \theta}g_{\theta\nu}
-\Gamma_{\nu\si}^{\ \ \theta}g_{\mu\theta}=0
\eeq

\beq\label{pm}
3.\ \ \ 
\na_\si g_{{\underset +\mu }{\underset -\nu}}=g_{\mu\nu},_\si-\Gamma_{\mu\si}^{\ \ \theta}g_{\theta\nu}
-\Gamma_{\si\nu}^{\ \ \theta}g_{\mu\theta}=0
\eeq

\beq\label{++}
4.\ \ \ 
\na_\si g_{{\underset +\mu }{\underset +\nu}}=g_{\mu\nu},_\si-\Gamma_{\mu\si}^{\ \ \theta}g_{\theta\nu}
-\Gamma_{\nu\si}^{\ \ \theta}g_{\mu\theta}=0
.\eeq

 Each gives rise to a relation between the metric tensor and the affine connection as follows.

\subsection{Case 1, - -, and Case 4, $++$} 
This corresponds to  $\na_\si g_{\underset -\mu \underset -\nu}=0$, and
 $\na_\si g_{{\underset +\mu }{\underset +\nu}}=0$.
By writing (\ref{mm}) with rotating indices in the usual way we find

\beq\label{affine}
\Gamma_{\mu\nu}^{\ \ \si}=S_{\mu\nu}^{\ \ \si}+
C_{\mu\nu}^{\ \ \si}+ \ga^{\si\theta}(S_{\theta\mu{\underset - \nu}}+S_{\theta\nu{\underset - \mu}})
\eeq
where

\beq\label{s}
S_{\theta\mu{\underset - \nu}}\equiv S_{\theta\mu}^{\ \ \la}\ga_{\la\nu}
\eeq

\no (the underbar should not be confused with the earlier definitions in (\ref{-}) and (\ref{+}))
and

\beq\label{c}
C_{\mu\nu}^{\ \ \si}=\frac12\ga^{\si\theta}(\ga_{\theta\mu,\nu}+\ga_{\theta\nu,\mu}-\ga_{\mu\nu,\theta})
.\eeq

\no We define
\beq
 \ph_{\mu\nu\si}\equiv3\ph_{[\mu\nu,\si]}\equiv
\ph_{\mu\nu,\si}+\ph_{\si\mu,\nu}+\ph_{\nu\si,\mu}
\eeq
and find

\beq
\frac12\ph_{\nu\mu\si}=S_{\mu\nu\si}+S_{\si\mu\nu}+S_{\nu\si\mu}
-(S_{\mu\nu{\underset -\si}}+S_{\si\mu{\underset -\nu}}+S_{\nu\si{\underset -\mu}})
.\eeq
A solution is given by

\beq\label{s1}
S_{\mu\nu}^{\ \ \si}=\frac16\ph^{\si\theta}\ph_{\mu\nu\theta}
.\eeq

\subsection{Case 2,  $ +-$ and Case = $-+$, }
This corresponds to $\na_\si g_{{\underset +\mu }{\underset -\nu}}=0$, and
 $\na_\si g_{{\underset -\mu }{\underset +\nu}}=0$. Performing the same kinds of manipulations as before we find,
 
\beq\label{affine2}
\Gamma_{\mu\nu}^{\ \ \si}=S_{\mu\nu}^{\ \ \si}+
C_{\mu\nu}^{\ \ \si}+ \ga^{\si\theta}(S_{\theta\mu}^{\ \ \la}\ph_{\la\nu}
+S_{\theta\nu}^{\ \ \la}\ph_{\la\mu})
\eeq
and

\beq
\frac12\ph_{\nu\mu\si}=S_{\mu\nu{\underset -\si}}+S_{\si\mu{\underset -\nu}}+S_{\nu\si{\underset -\mu}}
.\eeq
a solution of which is

\beq\label{s2}
S_{\mu\nu{\underset -\si}}=\frac12\ph_{\nu\mu\si}
.\eeq

\begin{widetext}
\hskip5em
\begin{table}[htp]
\caption{}
\begin{tabular}{|c|c|}\hline
 \multirow{3}{4em}{Case 1}$\na_\si g_{{\underset -\mu }{\underset -\nu}}=0$, and &  $\Gamma_{\mu\nu}^{\ \ \si}=S_{\mu\nu}^{\ \ \si}+
C_{\mu\nu}^{\ \ \si}+ \ga^{\si\theta}(S_{\theta\mu{\underset - \nu}}+S_{\theta\nu{\underset - \mu}})
$ \\
$\na_\si g_{{\underset +\mu }{\underset +\nu}}=0$  & $\frac12\ph_{\nu\mu\si}=S_{\mu\nu}^{\ \ \theta}\ph_{\theta\si}+S_{\si\mu}^{\ \ \theta}\ph_{\theta\nu}+S_{\nu\si}^{\ \ \theta}\ph_{\theta\mu}
$\\
   & $S_{\mu\nu}^{\ \ \si}=\frac16\ph^{\si\theta}\ph_{\mu\nu\theta}$\\
 \hline
 \multirow{2}{4em}{Case 2}$\na_\si g_{{\underset +\mu }{\underset -\nu}}=0$, and
& $\Gamma_{\mu\nu}^{\ \ \si}=S_{\mu\nu}^{\ \ \si}+
C_{\mu\nu}^{\ \ \si}+ \ga^{\si\theta}(S_{\theta\mu}^{\ \ \la}\ph_{\la\nu}
+S_{\theta\nu}^{\ \ \la}\ph_{\la\mu})$\\
 $\na_\si g_{{\underset -\mu }{\underset +\nu}}=0$ & 
 $\frac12\ph_{\nu\mu\si}=S_{\mu\nu{\underset -\si}}+S_{\si\mu{\underset -\nu}}+S_{\nu\si{\underset -\mu}}$
\\
 &$S_{\mu\nu{\underset -\si}}=\frac12\ph_{\nu\mu\si}$\\
 \hline
\end{tabular}
\label{default}
\end{table}
\end{widetext}

 From Case 1 above we see that as $\ph_{\mu\nu}$ goes to zero, the torsion is indeterminate. To see this explicitly we note that, for an antisymmetric $N x N$ quantity, $\ph^{\mu\nu}=1/(2\sqrt\ph)\ep^{\al\be\mu\nu}\ph_{\al\be}$ where $\ph$ is the determinant of $\ph_{\mu\nu}$ so that from (\ref{s1}) 

\beq
S_{\mu\nu}^{\ \  \si}=\frac{\ep^{\si\theta\al\be}\ph_{\al\be}\ph_{\mu\nu\theta}}
{12(\ph_{03}\ph_{12}-\ph_{02}\ph_{13}+\ph_{01}\ph_{23})}
.\eeq
This explicitly shows, as $\ph_{\mu\nu}\ra0$, the torsion is indeterminate. However, an interesting possibility is that the torsion goes to zero as $\ph_{\mu\nu}\ra $ constants. This gives empty space a chiral character and could only be meaningful if the universe has a net spin (or at least a net sense). This idea will not be pursued here.

From Case 2 and Case 3 we see that as $\ph_{\mu\nu}\ra0$ 
\beq\label{slim}
S_{\mu\nu\si}=\frac12\ph_{\mu\nu\si}
.\eeq
Thus, as $\ph_{\mu\nu}\ra0$ the torsion goes to zero. Not only that, we see this is exactly the same as (\ref{tordef1})
with $\ph_{\mu\nu}=(2/3)\psi_{\mu\nu}$.

To summarize the metricity condition, we adopt (\ref{q}) with $q_{\si\mu\nu}=-2S_{\si\mu\nu} $. This choice is based on the assumption that the torsion goes to zero in the limit the metric tensor becomes symmetric.
It is also noted  a term ${\cal F}_{\mu\nu}$ may be added to  $\ph_{\mu\nu}$ in (\ref{slim}) without changing the results if the skew symmetric derivative of
${\cal F}_{\mu\nu}$ vanishes, i.e., ${\cal F}_{[\mu\nu,\si]}= 0$.
\section{variational principle}

Although Einstein and others adopted the Palatini method,  there is no physical reason to do so. Here I will adopt the second order formalism for three reasons. First, it is the method used in all other field theories. In other words, once the canonical potential is identified, variations are always taken with respect to it and its derivatives. These theories are tried and true, and this serves as strong motivation to continue its use. The second reason is that the Palatini equations have been well studied. Einstein used them in an attempt to unify gravity and electromagnetism, but success was elusive. Moffat\cite{moffat} used the Palatini method but considered the antisymmetric part of the gravitational field to be part of gravity, and sought connections between that theory and rotation curves of galaxies (dark matter issues). Thus, between Einstein, Sciama, and Moffat, all the formulations are exhausted by the Palatini approach. The third reason is the most important: It is based on the physical idea that as the $\ph_{\mu\nu}$ goes to zero, the theory approaches the original 1915 version. This is very reasonable, and gives us the clue we need to establish the correct relation between the metric tensor and the affine connection.

To see all this explicitly let us start by, at first, considering
 $\mathscr{R}=\sqrt{-g}R$, and take the metric tensor and the affine connection to be independent, so we have

\beqa\label{vp1}
\de I=\de\int d^4x\mathscr{R}\ \ \ \ \ \\  \nonumber
=\int d^4x\left( \de(\sqrt{-g}g^{\al\be})R_{\al\be}+
\sqrt{-g}g^{\al\be}\de R_{\al\be}
\right)\\ \nonumber
=\int d^4x\left( -G^{\mu\nu}\de g_{\mu\nu}+
\tilde E^{\mu\nu}_{\ \ \si}\de \Gamma_{\mu \nu}^{\ \ \si}
\right) 
\eeqa
where
\no 
\beq\label{e}
\tilde E^{\mu\nu}_{\ \ \si}=\tilde g^{\al\be}\frac{\pa R_{\al\be}}{\pa \Gamma_{\mu \nu}^{\ \ \si}}
+
\tilde g^{\al\be}\frac{\pa R_{\al\be}}{\pa q_{\mu \nu}^{\ \ \si}}\frac{\pa q_{\mu\nu}^{\ \ \si}}{\pa \Gamma_{\mu \nu}^{\ \ \si}}
-
\pa_\eta\left(\tilde g^{\al\be}\frac{\pa R_{\al\be}}
{\pa \Gamma_{\mu \nu\ ,\eta}^{\ \ \si}}
\right)
\eeq
\no and where the tilde implies density, i.e.,  $\tilde E^{\mu\nu}_{\ \ \si}= \sqrt{-g}E^{\mu\nu}_{\ \ \si}$. Variations on the hyper-surface surrounding the volume are always taken to be zero.

Let us begin with the vacuum equations in which case $\de I=0$.
For the sake of comparison we note in passing the Palatini method gives

\beq\label{efirst}
 E^{\mu\nu}_{\  \ \si}=0
\eeq
and

\beq\label{gfirst}
G_{\mu\nu}=0
.\eeq

However, since we adopt the second order formalism we must choose the metricity condition to hold, and assume the Lagrangian density is a function of the metric tensor and its derivative. The affine connection is not an independent variable, so that we have

\beq\label{I}
\de I=
\int d^4x\sqrt{-g}\left(-G^{\mu\nu}  +L^{\mu\nu}
\right)\de g_{\mu\nu}
\eeq
where
\beq\label{L}
\tilde L^{\mu\nu}=
\tilde E^{\al\be}_{\ \ \ \si}\frac{\pa \Gamma_{\al\be }^{\ \ \si}}{\pa g_{\mu\nu}}
-
\pa_\theta\left(
\tilde E^{\al\be}_{\ \ \ \si}\frac{\pa \Gamma_{\al \be}^{\ \ \si}}{\pa g_{\mu\nu,\theta}}
\right)
\eeq
so that for the vacuum case we have

\beq\label{gsecond}
G^{\mu\nu}  +L^{\mu\nu}=0
.\eeq

The relation that follows from (\ref{e}) and the metricity condition is

\begin{widetext}
\beq
E^{\la\om}_{\ \ \ \si}=2S_\si^{ \ {\underset - \om} \lambda}-\na_\si g^{\lambda\om}
-\de_\si^\la q_\eta^{\ \om {\overset - \eta}}
-\frac12g^{\la\om}q_{\si\ \be}^{\ \be}-\frac12\de_\si^\la g^{\eta\om}q_{\et\ \be}^{\ \be}
+g^{\al\be}\left(S_{\be {\underset - \si}} ^{\ \ \ \om}\de_\al^\la
+S_{\al \si} ^{\ \ \ {\overset -\om}}\de_\be^\la
\right)
-S_{\overset-\si}^{\  {\overset-\la \om}}
.\eeq 
\end{widetext}
From (\ref{s2}) we can show the torsion trace, $S_\si$, is zero, which is not true in the Einstein variational principle. With this and (\ref{pm}) we have

\begin{widetext}
\beq\label{x}
L^{\mu\nu}=
\frac12\left(E^{\al\be\mu}S_{\al\be}^{\ \ \ \nu}+E^{\al\be\nu}S_{\al\be}^{\ \ \ \mu}
+\frac12q_{\et\ \si}^{\ \et}e^{\mu\nu\et}
\right)
+e^{\mu\nu\et}_{\ \ \ \ ;\et}
-\frac16 \left(f^{\mu\nu\et}_{\ \ \ \ ; \et}+\frac12f^{\mu\nu\et}
\right)
\eeq
\end{widetext}

where
\beq
e^{\mu\nu\et}=E^{\mu\et{\underset-\nu}}+E^{\et\mu{\underset-\nu}}-E^{\mu\nu{\underset-\et}}
\eeq
and
\beq
f^{\mu\nu\et}=E^{\mu\nu{\underset-\et}}+E^{\nu\et{\underset-\mu}}+E^{\et\mu{\underset-\nu}}
.\eeq

Thus, from the action

\beq
I= I_g+I_m,\ \ \ \de I=0
\eeq
where  $I_g$ is generalized to include the cosmological term $\Upsilon$, the field equations are found to be
\beq\label{nsfe}
G^{\mu\nu}-L^{\mu\nu}+\Upsilon g^{\mu\nu}\ =k T^{\mu\nu}
\eeq
where 

\beq
\de I_m\equiv 2k\int d^4x\sqrt{-g} T^{\mu\nu}\de g_{\mu\nu}
.\eeq

Now let us consider the matter action. In the torsion gravity theory with a symmetric metric tensor, as described above, it was shown that the energy momentum tensor is\cite{hamrev}

\beq
T^{\mu\nu}=\frac{\mu}{\sqrt{-g}}\int d^2\zeta\sqrt{-\la}\de(x-x(\zeta))x^\mu_{,a} x^\nu_{,b}
(\la^{ab}+\eta\ep^{ab})
\eeq
where $\la^{ab}$ is the two dimensional metric of the string and $\la$ is its determinant.
In this, $\ze_0$ is the timelike string coordinate and
$\ze_1$ is the spacelike coordinate, and in this section $a,b$
sum from 0 to 1. 
The string surface element is given by

\beq
d\si^{\mu\nu}=\ep^{ab}x^\mu_{,a}x^\nu_{,b}d^2\ze
\eeq
where
\beq
\sqrt{-\ga}\ep^{ab}=
\left(
\begin{array}{cc}
  0 & -1 \\
  1 & 0
\end{array}\right)
=
e^{ab}
.\eeq
This can be written in a naturally ``unified" form as follows. First we define

\beq\label{2dm}
\Lambda^{ab}={\cal \la}^{ab}+e^{ab}
\eeq
which is tantamount to taking the string metric to have an antisymmetric part. With this we can write

\beq
I_M=\mu\int d^2\zeta\sqrt{-\la}x^\mu_{,a}x^\nu_{,b}\Lambda^{ab}g_{\mu\nu}
\eeq
where here we assume $\la$ is not an explicit function of the metric tensor until after the variation. Thus, strings provide a very natural way to write the material action for a theory with a nonsymmetric metric tensor. However, it has been shown how to construct an action without strings. We will not dwell on this further here.

\section{Linearized theory}

Although this is a geometrical theory, we may think of the antisymmetric part of this as the equations for the spin and the symmetrical part as gravitation. This will be shown to be true in the linearized theory given below. To linearize we use (\ref{mt}) and assume the antisymmetric part is small compared to the symmetric part. In this we find

\beq
E^{\la\om}_{\ \ \ \si}=2S_\si^{\ \om\la}
\eeq
and

\beq
L^{\mu\nu}=2S^\mu_{\ \al\si}S^{\nu\al\si}
\eeq
and 

\beq
 G^{\mu\nu}=\ ^oG^{\mu\nu} +S^{\mu\nu\si}_{\ \ \ \ ;\si}-S^\mu_{\ \al\be}S^{\nu\al\be}
  \eeq
the  symmetrical part of the field equations become identical to (\ref{gfer}) and
 antisymmetric part of (\ref{nsfe}) is

\beq\label{tfe2}
\na_\si S_{\mu \nu}^{\ \ \si} -\frac12\ph_{\mu\nu}R=-kj_{\mu \nu}
\eeq
where again $j_{\mu\nu}\equiv (1/2)T_{[\mu\nu]}$.

This is certainly an noteworthy result. In the linearized theory it is the same as the torsional field equations with a symmetric metric tensor, i.e., equation (\ref{tfe2}) reduce to  (\ref{tfe}). However, if there is a region where $\ph$ is large, than the nonlinear terms may become exceedingly important, and may even represent a path to detecting torsion. 

It is interesting to note the effect of the cosmological term $\Upsilon$; the torsional field equations become (in the weak field limit and in vacuum),

\beq\label{tfe3}
\Box \ph^{\mu\nu} +\Upsilon \ph^{\mu\nu}=0
.\eeq
Thus, the cosmological constant finds a surprising connection, it gives the torsion quantum a mass!

In spacetime with a symmetric metric tensor and torsion, the field equations can always be broken into the Riemannian part and torsion terms. This cannot be done when the metric tensor is nonsymmetric, but we may proceed as follows. We write

\beq\label{affine3}
\Gamma_{\mu\nu}^{\ \ \si}=
C_{\mu\nu}^{\ \ \si}+\Lambda_{\mu\nu}^{\ \ \si}
 \eeq
where
\beq\label{ab1}
\Lambda_{\mu\nu}^{\ \ \si}=S_{\mu\nu}^{\ \ \si}+\ga^{\si\theta}(S_{\theta\mu}^{\ \ \la}\ph_{\la\nu}
+S_{\theta\nu}^{\ \ \la}\ph_{\la\mu})
\eeq
and
\beq
C_{\mu\nu}^{\ \ \si}=\frac12\ga^{\si\theta}\left(\ga_{\theta\mu,\nu}+\ga_{\theta\nu,\mu}-\ga_{\mu\nu,\theta}
\right)
.\eeq

We define the colon derivative as follows

\beq
A_{\al:\be}=A_{\al,\be} - C_{\al\be}^{\ \ \theta}A_\theta
\eeq

\no Using this in (\ref{ricci}) and (\ref{ab1}) we find

\beq
R_{\mu\nu}=r_{\mu\nu} +M_{\mu\nu}
\eeq
where

\beqa\label{ricci2}
r_{\mu\nu}
=C_{\mu\nu ,\si}^{\ \ \si}-C_{\si\nu ,\mu}^{\ \ \si}
+C_{\si\ph}^{\ \ \ \si}C_{\mu\nu}^{\ \ \ \ph}
-C_{\mu\ph}^{\ \ \ \si}C_{\si\nu}^{\ \ \ \ph}\ \ \ \ \ 
\eeqa
and where

\beq\label{ricci3}
M_{\mu\nu}=\Lambda_{\mu\nu\ :\si}^{\ \  \ \si}
-\Lambda_{\si\nu\ :\mu}^{\ \  \ \si}
+\Lambda_{\si\theta}^{\ \ \si}\Lambda_{\mu\nu}^{\ \ \theta}
-\Lambda_{\mu\theta}^{\ \ \si}\Lambda_{\si\nu}^{\ \ \theta}
.\eeq
Thus, in the limit $\ph_{\mu\nu}=0$, $M_{\mu\nu}=0$, $g_{\mu\nu} $ becomes the symmetric metric tensor and $C_{\mu\nu}$ is the Ricci tensor of Riemannian space-time. 

In the linearized (in $\ph_{\mu\nu}$) theory we use

\beqa
g_{\mu\nu}=\ga_{\mu\nu} +\ph_{\mu\nu}\\ \nonumber
g^{\mu\nu}=\ga^{\mu\nu}-\ga^{\nu\al}\ga^{\mu\be} \ph_{\al\be}
\eeqa
and find

\beq\label{s3}
S_{\mu\nu\si}=\frac12\ph_{\nu\mu\si}
\eeq
which is the same as (\ref{tordef1}) (to within a numerical factor).
Also, the curvature scalar becomes,

\beq
R=C-S_{\mu\nu\si}S^{\mu\nu\si}
\eeq
where $C=C_{\mu\nu}\ga^{\mu\nu}$. This is precisely the same as (\ref{rplus}), since $R$
becomes the Riemannian Ricci tensor in this limit.

Thus, many of the results of the theory with a symmetric metric tensor may be borrowed, since the equations, in the weak field limit, are the same. For example, the Minkowski limit solution to the torsional field equations were found for a static source. to see this, the torsion dual is defined as

\begin{equation}\label{bmu}
b_{\mu} = \epsilon_{\mu \alpha \beta \gamma}S^{\alpha \beta \gamma}
\end{equation}
and a vacuum solution to the torsional field equations are
\begin{equation}\label{torsol}
\bm{b} =  \frac{3k}{4\pi}{S\over r^{3}}
\left(2 \cos(\theta) \hat r + \sin(\theta)\hat \theta \right)
.\end{equation}

Thus, the linearized theory agrees precisely with the theory of spin gravity described above.
This not only affirms the motivation, but more importantly, we can rely on the body of work already done on that theory. So we know the origin of the torsion field is intrinsic spin, and the correct law for conservation of total angular momentum results from this interpretation.\cite{hamrev} Bounds on the coupling constant were examined, and coupling to the Dirac equation was accomplished.\cite{hamprd} The details of its connections to string theory were studied in detail, and it was shown spin flipping generates torsion waves, and the power was calculated.\cite{hamtp} All of these apply here, in the linearized theory, however the full nonlinear theory brings new interactions to the table that should be examined further.

It may seem strange that a theory with an nonsymmetric metric tensor can reduce to one with a symmetric metric tensor and torsion. To see this in general we note, for the determinants, $g$, $\ga$, and $\ph$ 
of $g_{\mu\nu}$, $\ga_{\mu\nu}$, and $\ph_{\mu\nu}$,
\beq\label{det}
g=\ga+\ph +\frac\gamma2\ga^{\al\be}\ga^{\mu\nu}\ph_{\al\mu}\ph_{\be\nu}
.\eeq
In the weak field limit

\beq\label{ab}
\Lambda_{\mu\nu}^{\ \ \si}=S_{\mu\nu}^{\ \ \si}
\eeq
and the torsion is given by (\ref{slim}). Putting this together, to lowest order we find

\beq
I=\int d^4x\sqrt{-g}R
\ra \int d^4x\sqrt{-\ga}(1+\frac{\ph}{2\ga})(C- S_{\mu\nu \si}S^{\mu\nu \si})
.\eeq

In the linearized theory $C$ reduces to the Riemannian $R$ and $\ga_{\mu\nu}$ becomes the symmetric $g_{\mu\nu}$ of Riemannian spacetime. Thus, the theory with a nonsymmetric metric tensor reduces to a theory with a symmetric metric tensor and totally antisymmetric torsion so long as $\ph/2\ga<<1$.

\section{enter electromagnetism}

We began by abandoning all ties to electromagnetism and showed the antisymmetric part of the metric tensor is related to spin. It was shown in the weak field limit equations reduce to those of torsion gravity with a symmetric metric tensor. In that theory the spin potential possessed the gauge invariance $\ps_{\mu\nu}\ra \ps_{\mu\nu}+\xi_{[\mu,\nu]}$. However, the metric tensor enjoys no such freedom.

This also happens in string theory when a point charge is put on a brane. To maintain gauge invariance a term is added to $\ps_{\mu\nu}$. We adopt the same notion here and let

\beq\label{gi}
g_{[\mu\nu] } = \ph_{\mu\nu} +{\cal F}_{\mu\nu}
\eeq
where
\beq\label{fdef}
{\cal F}_{\mu\nu}\equiv \xi_{[\nu,\mu]}
.\eeq
With  this, gauge invariance is restored for the combined gauge transformation 
\beq\label{gi2}
\ph_{\mu\nu} \ra \ph_{\mu\nu} +\la_{[\mu,\nu]}
\eeq
and
\beq
\xi_\mu\ra \xi_\mu+\la_\mu
.\eeq
It was noted after (\ref{slim}) the term ${\cal F}_{\mu\nu}$ may be added to the solution we found for the torsion provided 

\beq\label{hom}
{\cal F}_{[\mu\nu,\si]}+{\cal F}_{[\si\mu,\nu]}+{\cal F}_{[\nu\si,\mu]}= 0
.\eeq

In light of (\ref{fdef}) and (\ref{hom}),
it is natural to ask if $F_{\mu\nu} =\ka{\cal F}_{\mu\nu}$ is the electromagnetic field (the constant $\ka$ must be included for dimensional reasons, and we may consider the potential to be $A_\mu=\ka\xi_\mu)$. 
If so, there would have to be the kinetic term in the Lagrangian.
It turns out that term is there. Expanding
$\sqrt{-g}$ in the weak field limit, from  the last term in (\ref{det}) we see a term $F_{\mu\nu}F^{\mu\nu}$ is already present. In fact, the Lagrangian becomes, in the weal field limit,

\beqa
\int\sqrt{-g}d^4(R+\Upsilon)
\ra \ \ \ \ \ \ \ \ \ \ \ \ \ \ \ \ \ \ \ \ \ \ \ \\ \nonumber
\int\sqrt{-\ga}d^4x\left( ^oR-S^{\al\be\si}S_{\al\be\si}+\frac{1}{16\pi}F^{\al\be}F_{\al\be}
\right)
\eeqa
where $\ka$ is taken to be $\ka=G/c^2\sqrt{\Upsilon}$.
Thus, gravitation with a nonsymmetric metric tensor reduces to, in the weal field limit, the theory of gravitation with spin and the electromagnetic field. 

It is also interesting to note the cosmological constant must be non-zero if the electromagnetic field is present. Another way of stating this is, the cosmological constant must be non-zero if the theory is gauge invariant, so there is finally a strong theoretical reason for the existence of the cosmological constant. It is also worth noting, in cgs and using the currently accepted value of the cosmological constant, $\ka$ is near unity.

\section{discussion}

The general theory of relativity of 1915 (GR) has been tested to be  successful and, for that reason, the theory presented here follows general relativity as closely as possible with only one change, initially, being that the metric tensor is not symmetric. It was explained above, due to spin, the metric tensor should not  be symmetric, and it gives rise to a non-vanishing torsion field. In this paper the second order formalism was used, necessitating the need of a connection  between the metric tensor and the affine connection. It was assumed that as the nonsymmetric part of the metric tensor goes to zero, the torsion should go to zero. This fixed the value of the non-metricity.

The linearized equations were presented and it was found the theory reduces to a previous theory in which the metric was symmetric and the torsion was totally antisymmetric. In that theory the equations of motion, interactions, and bounds on coupling constants were all worked out, so we may borrow them here. However, new nonlinear and gravitational interactions are predicted and need investigation.

In order to formulate the material action it is shown a natural choice is the string action. The two dimensional string metric was taken to be nonsymmetric according to (\ref{2dm}). However this can be generalized further by letting $e_{ab}\ra\et e_{ab}$ where $\eta$ is an undetermined coupling constant.

Although theories with non-vanishing torsion have been studied for a century now, an argument sometimes used to dismiss them is that there is no reason to assume torsion exists in the first place, and if it does, there is no reason to assume it is given by (\ref{tordef1}). But with a nonsymmetric metric tensor it must exist, and must be given by
(\ref{tordef1}), in the weak field limit.

It was also shown if a cosmological constant is introduced, it gives rise to a mass of the torsion quanta, and gauge invariance is lost even in the linearized theory. But even without the mass term gauge invariance of form $\ph_{\mu\nu}\ra\ph_{\mu\nu}+\xi_{[\mu,\nu]}$ is lost in the general theory. This is obvious since the metric tensor enjoys no such invariance. The same thing happens in string theory for open strings. In that case, the electromagnetic field is introduced to resurrect gauge invariance. This is an exciting result, leading to a modern interpretation of a connection between gravitation and electromagnetism.

\section{appendices}

For completeness we derive the field equations for the case of zero metricity.
In this case (\ref{L}) and (\ref{gsecond}) still hold, but $ E^{\la\om}_{\ \ \si}$ is different. In fact it is,

\begin{widetext}
\beq
 E^{\la\om}_{\ \ \si}=g^{\al\be}\Ga_{\al\be}^{\ \ \ \om}\de^\la_\si+g^{\la\om}\Ga_{\ga\si}^{\ \ \ \ga}
 -g^{\la\be}\Ga_{\si\be}^{\ \ \ \om}-g^{\al\om}\Ga_{\al\si}^{\ \ \ \la}
 -\left( \frac{\sqrt{-g},_\si}{\sqrt{-g}}g^{\la\om}+g^{\la\om},_\si\right)
 +\de^\la_\et\left( \frac{\sqrt{-g},_\si}{\sqrt{-g}}g^{\eta\om}+g^{\eta\om},_\et\right)
 \eeq
\end{widetext}
which, with (\ref{mm}) set equal to zero reduces to
\beq
 E^{\la\om}_{\ \ \si}=2\de^\la_\si g^{\et\om}S_\et+2g^{\theta\om}S_{\si\theta}^{\ \ \la}
 -2g^{\la\om}S_\si
\eeq
where $S_\et=S_{\et\si}^{\ \ \si}$ does not vanish in this case. With this we can work out (\ref{L}). From (\ref{I}) we may note

\beq
L^{\mu\nu}\de g_{\mu\nu}=L^{\mu\nu}(\de \ga_{\mu\nu}+\de \ph_{\mu\nu})
\eeq
so we can work out the symmetric part, $L_{S}^{\mu\nu}$ and then the antisymmetric part
$L_{A}^{\mu\nu}$. Thus, assuming taking the symmetric part (in $\mu\nu$) is implied,

\begin{widetext}
\beqa\label{s33}
\tilde L_{S}^{\mu\nu}= \tilde E^{\la\om}_{\ \ \si}\left[
-\ga^{\mu\si}C_{\la\om}^{\ \ \nu}
-\ga^{\si\mu}S_{\la\ \ {\underset - \om}}^{\ \ {\underset - \nu}}
+\de^\nu_\om S_{\la}^{\  {\underset - \si}\mu}
+\de^\nu_\la S_\om^{\  {\underset - \si }\mu}
-
\ga^{\mu\si}S_{\om\   {\underset-\la }}^{\  {\underset - \nu}}
\right]\\ 
\nonumber
-\frac{\pa_\et}{2}\left[ \tilde E^{\la\om}_{\ \ \si}\left(\ga^{\si\mu}\de^{\nu\et}_{\la\om}
+\ga^{\si\mu}\de^{\nu\et}_{\om\la}
-\ga^{\si\et}\de^{\mu\nu}_{\la\om}
\right)\right]
\eeqa
\end{widetext}
we adopt the unconventional notation $\de^{\nu\et}_{\la\om}=\de^\nu_\la\de^\et_\om$.
Defining

\beq
e^{\nu\et{\underset-\mu}}\equiv E^{\nu\et{\underset-\mu}}
+E^{\et\nu{\underset-\mu}}
-E^{\mu\nu{\underset-\et}}
\eeq
(\ref{s333}) may be written as

\begin{widetext}
\beq\label{s4}
L_{S}^{\mu\nu}=
- E^{\la\om{\underset-\mu}} S_{\la\om}^{\  \ \nu}
+E^{\la\nu}_{\ \ \om} S_\la^{\  {\underset - \om }\mu}
-S_\et e^{\nu\et{\underset-\mu}} 
-\frac12\na_\et e^{\et\nu{\underset-\mu}} 
.\eeq
\end{widetext}

Now we may find the nonsymmetric part $L_{A}^{\mu\nu}$, where the antisymmetric part is implied. It is,

\begin{widetext}
\beqa	
L_{A}^{\mu\nu}=E^{\la\om}_{\ \ \si}(
-\ph^{\si\nu} S_{\la\om}^{\ \ \mu}
-\ga^{\si\theta}\ga_{\xi\om}\ph^{\xi\nu}S_{\la\theta}^{\ \ \mu}
-\ga^{\si\theta}\ga_{\xi\la}\ph^{\xi\nu}S_{\om\theta}^{\ \ \mu}
)
-\frac{\pa_\et}{6}[\tilde E^{[\la\om]}_{\ \ \ \ \si}
(
\ph^{\si\et}\de^{\mu\nu}_{\la\om}
+2\ph^{\si\mu}\de^{\nu\et}_{\la\om}
)
]
\\ \nonumber
+2(\ga^{\si\nu}\ga_{\xi\ph}\ph^{\xi\et}\de^\mu_\om
+\ga^{\si\et}\ga_{\xi\om}\ph^{\xi\mu}\de^\nu_\la
+\ga^{\si\mu}\ga_{\xi\om}\ph^{\xi\nu}\de^\et_\la
)
\eeqa
\end{widetext}
which  may be written

\begin{widetext}
\beqa	
L_{A}^{\mu\nu}=E^{\la\om}_{\ \ \si}\ph^{\si\nu}S_{\la\om}^{\ \ \mu}
-E^{\la\om{\underset-\theta}}
(\ga_{\xi\om}\ph^{\xi\nu}S_{\la\theta}^{\ \ \mu} +\ga_{\xi\la}\ph^{\xi\nu}S_{\om\theta}^{\ \ \mu})
\\ \nonumber
-\frac16\left[(\na_\et+S_\et)(E^{\mu\nu}_{\ \ \si}\ph^{\si\et}
+E^{\nu\et}_{\ \ \si}\ph^{\si\mu}
-E^{\et\nu}_{\ \ \si}\ph^{\si\mu})
+2S_{\theta\et}^{\ \ \mu}(E^{\theta\nu}_{\ \ \si}\ph^{\si\et}+E^{\nu\theta}_{\ \ \si}\ph^{\si\et}
\right]
\eeqa
.\end{widetext}
These, with (\ref{gsecond}), are the field equations for minus minus metricity.

\ed
\begin{thebibliography}{}


\bibitem{sciama} D. W. Sciama, Proc. Cambridge Philos. Soc., {\bf 54}, 72 (1958). 

\bibitem{bel} F. Belinfante, Physica {\bf 6}, 887 (1939).



\bibitem{pap} A. Papapetrou  Philosophical Magazine Series 7, 40:308, p. 237, (1949); and Ref. \cite{bel}.

\bibitem{tg}  R. T. Hammond, 
Grav. {\bf 26}, 247 (1994).



\bibitem{st} M. Kalb and Pierre Ramond, Phys. Rev. D {\bf 9}, 2273 (1974).

\bibitem{e1} A. Einstein, Ann. Math. {\bf 46}, 578 (1945).

\bibitem{e2} A. Einstein and E. G. Straus, Ann. Math. {\bf 47}, 731 (1946).

\bibitem{e3} A. Einstein, Rev. Mod. Phys. {\bf 20}, 35 (1948).

\bibitem{goen} H. F. M. Goenner, Living Rev. Relativ. (2014) 17: 5. https://doi.org/10.12942/lrr-2014-5


\bibitem{s} E. Schr\"odinger, Proc. R. Ir. Acad. A, {\bf 49 }, 43 (1943);
{\bf 49}, 237 (1944);
{\bf 49}, (1944);
{\bf 51}, 41 (1946).


 \bibitem{hehl} F. W. Hehl, P. von der Heyde, G. D. Kerlick, and J. M. Nester
 {\it Rev. Mod. Phys.} {\bf 48} , 393 (1976).
 
 
\bibitem{hamrev} R. T. Hammond,  Rep.  Prog. Phys., {\bf 65}, 599 (2002).

\bibitem{shap} I. Shapiro, Phys. Rep. {\bf 357}, 113 (2002).

\bibitem{moffat} J. W. Moffat, Phys. Rev. D {\bf 19}, 3554 (1979).

\bibitem{dam} T. Damour, S. Deser, and J. McCarthy
Phys. Rev. D {bf 47}, 1541 (1993).

\bibitem{ham1} R. T.  Hammond,  Int. Journal Mod. Phys. D, {\bf 22}, 1342009 (2013).
 
\bibitem{mc}  B. T.  McInnes, Journal of Physics A {\bf13}, 3657 (1980).

\bibitem{yildez} A. Yildiz, M. K.  Hinders, B. A. Rhodes,  et al. Nuov. Cim. A {\bf 102}, 1417 (1989).


\bibitem{hamx} R. T. Hammond, 
Int. J. Mod. Phys. Lett. D {\bf 27},  1847005 (2018). 

\bibitem{ham2} R. T. Hammond, Class. Quantum Grav. {\bf  12}, 279 (1995).


\bibitem{ton} M. A. Tonnelat, {\it Einstein's Theory of Unified Fields} (Gordon \& Breach, 1965).


\bibitem{hamprd}  R. T. Hammond,
Phys. Rev. D15, {\bf 52}, 6918 (1995).



\bibitem{hamtp} R. T. Hammond,
Gen. Rel.  Grav. {\bf 29}, 727 (1997).






\end{thebibliography}
